# Automated Model-Free Sorting of Single-Molecule Fluorescence Events Using a Deep Learning Based Hidden-State Model

Wenqi Zeng [2]\*, Shuqi Zhou [1]\*, Yuan Yao [2✉], Chunlai Chen [1✉]

[1] State Key Laboratory of Membrane Biology, Beijing Frontier Research Center for Biological Structure, School of Life Sciences, Tsinghua University, Beijing 100084, China.

[2] Department of Mathematics, the Hong Kong University of Science and Technology.

\* These authors contributed equally: Wenqi Zeng, Shuqi Zhou

✉ Correspondence: yuany@ust.hk, chunlai@mail.tsinghua.edu.cn

## Abstract

Single-molecule fluorescence assays enable high-resolution analysis of biomolecular dynamics, but traditional analysis pipelines are labor-intensive and rely on users' experience, limiting scalability and reproducibility. Recent deep learning models have automated aspects of data processing, yet many still require manual thresholds, complex architectures, or extensive labeled data. Therefore, we present DASH, a fully streamlined architecture for trace classification, state assignment, and automatic sorting that requires no user input. DASH demonstrates robust performance across users and experimental conditions both in equilibrium and non-equilibrium systems such as Cas12a-mediated DNA cleavage. This paper proposes a novel strategy for the automatic and detailed sorting of single-molecule fluorescence events. The dynamic cleavage process of Cas12a is used as an example to provide a comprehensive analysis. This approach is crucial for studying biokinetic structural changes at the single-molecule level.

## Introduction

Single-molecule fluorescence resonance energy transfer (FRET) is a leading technique for the study of biomolecular dynamics and interactions at the nanoscale. In the FRET process, energy is transferred[1,2] between two fluorophores (light-emitting molecules) when they are close enough together (in nanometers)[3]. This process depends on how close the fluorophores are. By overcoming the limitations of traditional ensemble FRET measurements, which often obscure critical molecular details due to population averaging, single-molecule FRET provides unprecedented insights into conformational changes and interaction dynamics. It also enables researchers to observe real-time structural dynamics[4] and interactions that remain elusive in ensemble measurements. The versatility of single-molecule FRET is evident across various applications in biophysics and molecular biology, including the study of conformational dynamics of proteins and nucleic acids [4,5], such as the ribosome[6–9], Cas proteins[10–13], membrane proteins[14–18], chromatin[19], RNA riboswitches[20,21], elucidation of biomolecular interactions[22].

A typical single-molecule FRET data analysis involves region alignment, point spread function identification, fluorescence intensities of donor and acceptor fluorophores measurement, trace classification, and state assignment for subsequent kinetic quantification. Extracting kinetic parameters and recognizing distinct FRET states from single-molecule FRET data can be challenging due to experimental limitations and the inherent complexity of biomolecular dynamics. Current methods developed to address these challenges can be broadly categorized into Hidden Markov Model (HMM)-based approaches[23–26] and deep learning approaches[27–29]. HMMs are employed to infer the state-to-state transitions s, which is robust for noisy data and complex kinetic schemes, but HMMs assume that single-molecule events occur on slower timescales than those of data acquisition, which may not always be true, and always required input number of states. Deep learning models such as Convolutional Neural Network (CNN)[30] and Long Short-Term Memory (LSTM)[31] have become pervasive. These models accelerate single-molecule fluorescence events analysis by handling large and complex datasets effectively and automatically. However, it should be noted that current analytical methods all rely to a certain extent on some input or judgment by users during the analysis process, which can lead to a biased selection of single-molecule fluorescence events and affect the conclusions drawn.

To address these limitations, we introduce DASH (Deep learning-based Automatic Sorting by identifying Hidden states in fluorescence events), a deep learning framework designed to automatically sort fluorescence events while eliminating user-dependent biases. DASH integrates two modules: trace classification and state assignment of single-molecule fluorescence events, offering a simplified architecture that reduces model complexity and training overhead. Importantly, DASH segments events and assigns discrete FRET states without the need for user-defined thresholds or manual specification of the number of states. This fully automated workflow minimizes user dependence and enhances reproducibility across different expertise levels, helping the user to have a more comprehensive understanding of the experimental system in the absence of a priori knowledge. Here, we validate the performance of DASH through the use of three equilibrium systems (Ribosome[32], $\beta_2$ adrenergic receptor[33] and Holliday junction[34]) and two non-equilibrium systems (Cas12a[35] and CasX (also known as Cas12e)[36]), the results of further analysis based on trace classification and state assignment results all show a high consistency with existing knowledge.

## Results

### DASH training and validation

DASH can be divided into two stages, trace classification, and state assignment (Figure 1a).

**Trace classification**: The model for trace classification is trained based on five datasets from three distinct biological samples: the ribosome[32], the $\beta_2$ adrenergic receptor[33], and the DNA Holliday Junction[34]. These include triple-classification labels, which were manually annotated by human experts and DASH, please see our previous study for details of classification rules[37]. The trace classification adopts bi-LSTM (Bidirectional Long Short-Term Memory) network and is equipped with CRF (conditional random field). Bi-LSTM networks are well-suited for handling temporal dependencies in sequential data, such as FRET traces. They can capture both forward and backward dependencies in an auto-aggressive manner, which can effectively process the time-ordered nature of FRET traces, distinguishing between different states based on intensity fluctuations. CRF layers are particularly applied to utilize the embeddings from bi-LSTM and modeling state transitions and dependencies between labels. In FRET analysis, CRFs can help identify the sequence of molecular states by considering the probabilities of transitioning between these states. By optimizing the transition matrix and calculating the whole sentence likelihood, CRFs ensure that the predicted state sequence is globally optimal, considering the dependencies between adjacent states. This is essential for accurately capturing kinetic information from FRET traces.

The combination of bi-LSTM and CRF is to use bi-LSTM networks to extract robust features from the intensity traces and use CRFs to further refine these features by imposing constraints on the sequence of states, helping to distinguish true state transitions from noise-induced fluctuations. For this stage, the input is the raw two-color trace signals from single-molecule two-color fluorescence experiments, the output is a three-class category, 'donor photobleached', 'acceptor photobleached', and 'discarded'. (Figure 1b, Supplementary Figure S1).

Compared to manually classified results, Deep HSM achieves an average F1 score of 0.91 (Figure 1c) on five datasets, the model, indicating that the model performs comparably to that of human experts. The consistency between classifications from two independent experts was approximately 0.9[37]. Furthermore, the high consistency of the identification of FRET events demonstrates the model's suitability for determining the appropriate segmentation timepoint (Figure 1d).

**State Assignment**: As for the second stage, considering the lack of a priori knowledge of most of the experimental data, we choose to use the HMM to assign discrete states for trace segments obtained in stage 1. The assumption in HMM naturally aligns with FRET traces which capture time-dependent molecular conformational changes: Discrete hidden states correspond to distinct molecular configurations (low/mid/high FRET efficiencies). Observed emissions (fluorescence intensities) depend probabilistically on these hidden states. Markovian transitions between states model kinetics (e.g., binding/unbinding events or structural rearrangements). The FRET of single experiments is concated and estimated by HMM to determine the optimal number of discrete states, which is decided by a weighted score of BIC and silhouette score. Only FRET traces classified as 'donor photobleached' or 'acceptor photobleached' proceed to the second stage of analysis. HMMs are employed to model the observed fluorescence intensities as emissions from discrete hidden states, enabling the identification of distinct FRET states. The output of this stage includes inferred state trajectories, which represent the most likely sequence of hidden states over time. Additionally, frequent events are identified through frequency counts of state transitions, providing insights into the dynamics of molecular interactions. This approach ensures state assignment and facilitates the extraction of key kinetic and structural information from single-molecule FRET traces.

**Accurate FRET states assignment in steady two-color traces.**

Based on the results from DASH's state assignment for the five datasets, we can obtain the FRET distribution results for different states. The model demonstrates the ability to accurately assign FRET states that are very close to each other, as well as those that account for a small percentage of the total (Figure 2a). The FRET distribution plots based on the DASH's trace classification and state assignment results are consistent with the results of our previous paper[37], which also proves the accuracy of DASH in trace classification and state assignment. Furthermore, traces can be finely sorted automatically according to distinct FRET patterns (see categorized FRET distributions and traces in Supplementary Figures S2-S6). Here, we define a pattern as the occurrence of a FRET state transfer within the current trace and the direction of that state transfer, which considers both the number of states and the state-to-state transitions direction.

Additionally, based on the state assignment results, we can quantify the proportion of state transition times across all traces (Figure 2b). These results provide users with a preliminary understanding of whether the experimental data exhibit static FRET or transient FRET. For example, the Holliday Junction shows a transition between two FRET states under high salt conditions (500 mM $Mg^{2+}$). Conversely, under low salt conditions (1 mM $Mg^{2+}$), the dynamics are accelerated beyond the camera's set time resolution, leading to an average representation of the two states and resulting in most traces exhibiting one FRET state. For transient traces, the state-to-state transition direction and probability show the dynamic changes of traces (Figure 2c). Please refer to Supplementary Tables 1-5 for all patterns sorted automatically after the DASH analysis. In summary, the automatic sorting method offers an intuitive understanding of the FRET states, the percentage of different states, and the transitions between states.

**Automatic sorting based on DASH classification following the state assignment of AsCas12a.**

DASH's state assignment-based automatic clustering enhances its performance in analyzing non-equilibrium dynamic systems, particularly in the context of single-molecule FRET experiments where fluorescence signals appear during data acquisition and exhibit transient states. Non-equilibrium dynamics are inherent in many biological processes, including ribosome initiation and elongation[32], $\beta_2$-adrenergic receptor signaling[33], and Cas protein target recognition and cleavage[13,35,36]. Analyzing these processes demands tools capable of capturing fluorescence events, inferring state transitions, and detecting FRET time-dependent changes in population distributions.

In this study, we reanalyze data from previously published *Acidaminococcus sp.* Cas12a (AsCas12a) recognition and cleavage design[35,37]. AsCas12a can be guided by crRNA to recognize DNA target sites adjacent to PAM sites to initiate R-loop formation and double-stranded DNA (dsDNA) cleavage. Utilizing single-molecule FRET, we observed the interactions between Cy3-labeled Cas12a/crRNA complexes and immobilized Cy5-labeled DNA, allowing to sequentially capture the formation of the Cas12a/crRNA/DNA complex, non-target strand (NTS) cleavage, target strand (TS) cleavage, and DNA fragment release (Figure 3a).

To illustrate the DASH-based data analysis process, a case study is presented using the AsCas12a cleavage assay with a fully matched DNA target. In the trace classification stage, DASH classifies long-term intensity-frame fluorescence traces and selected fluorescence event segments, accurately determining their start and end time points (Figure 3b). Subsequently, the selected segments for "donor photobleaching" and "acceptor photobleaching" are automatically processed in stage 2 for FRET state assignment. Based on the state assignment results of all selected segments, we generated time-dependent

FRET probability density plots (Figure 3c) by synchronizing all identified events at the initiation or termination time points of the FRET signals. From these plots, As0MM exhibits a total of five states, with States 1 to 4 (S1-S4) sequentially corresponding to the initial R-loop formation, NTS cleavage, and TS cleavage. Notably, State 0 (S0) has a FRET value close to 0, indicating that the Cy3-labeled Cas12a/crRNA complex searches for immobilized Cy5-labeled target DNA through one-dimensional (1D) diffusion. This process is typically short (less than 5 frames) but frequent, suggesting that DASH effectively manages complex and transient FRET states. State 2 (S2) occupies the majority of the FRET distribution, indicating that most of the Cas12a remains at the R-loop formation step or stays at this step for an extended period (Figure 3c). The FRET distribution plots based on the DASH classification and state assignment results are consistent with the findings of our previous paper[35], further validating the accuracy of DASH in trace classification and state assignment.

To further understand the various types of the Cas12a cleavage process, we analyzed FRET probability density plots over time and the corresponding FRET distributions based on the number of states present within a fluorescence event. (Figure 3d). Our analysis showed that the majority of fluorescence events (about 35% of the total) remained in a single state, while only 246 fluorescence events (2.48%) out of a total of 3,767 selected events were observed across all four cleavage steps of AsCas12a (Figure 3d). Each category with a different number of states can be further categorized by a specific pattern that takes into account both the number of states and the state-to-state transitions direction (Figures 3e-3g). All fluorescence events can be categorized first by transition number and then by pattern (Figure 3h). All patterns automatically categorized after the DASH analysis are shown in Supplementary Table 6. This model-free sorting method facilitates quick access for users to the individual categories representing different biological activities.

Known for its widespread use in genome editing, AsCas12a's activity is affected by mismatches. We attempted to introduce mismatches between the crRNA and target DNA for single-molecule FRET assays, utilizing DASH for trace classification, state assignment, and automatic sorting. As expected, the introduction of mismatches hinders the transition to high FRET cleavage (Figures 4a, 4d, 4g and Supplementary Figure 7) and reduces the proportion of more than four transition times (Figures 4b, 4e, 4h). From the subcategorization based on number of states present within an event ((Figures 4c, 4f, 4i), the events containing all four cleavage steps of AsCas12a decreased to 1.27% (3MM), 2.94% (6MM), and 0.20% (7MM). Each category of different number of states within a fluorescence event can be further sorted by specific pattern (Supplementary Figures 8-10), all patterns automatically categorized after the DASH analysis are shown in Supplementary Tables 7-9.

In summary, the state assignment results based on DASH provide an effective automatic sorting method that enables the identification of a relatively small group of fluorescence events within the majority, without being obscured by the average results presented by the whole dataset.

**Automatic sorting based on DASH classification following the state assignment of CasX (Cas12e)**

CasX, also known as Cas12e, is a class 2 CRISPR-Cas system that is smaller in size[38] compared to the widely used Cas9 and Cas12a, showing considerable promise in genome editing. It specifically recognizes the 5'-TTCN PAM and demonstrates significant genome editing efficacy when combined with single guide RNA (sgRNA) in human cells. Two homologous proteins, *Deltaproteobacteria* (DpbCasX) and *Plantomycetes* (PlmCasX), have been extensively studied[39]. In this study, we collected data from previously published CasX recognition and cleavage designs[36], where Cy3 and Cy5 fluorophores were attached to the 3' end of sgRNA and the target strand, 34 nucleotides away from the PAM distal end. In the trace classification stage, DASH classifies long-term intensity-frame fluorescence traces and selected fluorescence events, accurately determining their start and end time points (Figures 5a,5d). Subsequently, the selected segments for "donor photobleaching" and "acceptor photobleaching" are automatically processed in stage 2 for FRET state assignment. Based on the state assignment results of all selected fluorescence events, three different conformational states of dpbCasX during cleavage are revealed: R-loop formation (corresponding to State 1), NTS pre-cleavage (corresponding to State 2), and TS pre-cleavage (corresponding to State 3) (Figures 5b-5c and Supplementary Figure 11a). In contrast, plmCasX exhibits only two FRET states due to different cleavage steps on the NTS (Figures 5e-5f and Supplementary Figure 11b).

The time-dependent FRET probability density plots obtained through DASH trace classification and state assignment align with previous studies[36] but with a greater number of identified fluorescence events. Through automatic sorting analysis, we found that the majority of fluorescence events remain at the R-loop formation step. Productive binding can be effectively separated from all events among numerous datasets, showing consistent time-dependent FRET changes in line with previous findings. Each category of different number of states within a fluorescence event can be further sorted by specific pattern (Supplementary Figures 12-13), all patterns automatically categorized after the DASH analysis are shown in Supplementary Tables 10-11.

## Discussion

In summary, DASH is a deep learning-based framework for identifying single-molecule fluorescence events by integrating trace classification with state assignment. The method segments FRET traces into three categories and assign discrete states to selected segments, enabling the detection of diverse transition patterns. It effectively captures both steady and dynamically emerging events in two-color traces of varying lengths without relying on any user-defined thresholds. To maximize accessibility, we provide open-source Python code for model training, trace classification (emerging events identification), and state assignment. Based on the results of DASH's trace classification and state assignment, we can automatically generate a further sorting of fluorescence events, which can be refined into specific patterns. This refinement takes into account the FRET states exhibited at the single-molecule level, reflecting the number of conformational changes as well as the directed transitions between different FRET states. The advantage of this automatic sorting method, grounded in state assignment, is that it does not require the experimenter to possess a priori knowledge of the experimental system. Instead, it allows for the analysis of each pattern-based category and resolves conclusions or trends that may be obscured by massive amounts of data. This approach effectively demonstrates how experimental data can inform and guide the interpretation of results, enhancing the understanding of the underlying biological processes. This capability not only streamlines the analysis process but also provides a more objective framework for interpreting complex fluorescence data.

## Methods

### Training datasets

In consideration of the potential variance in fluorescence intensity levels due to different application systems, five single-molecule, two-color FRET datasets were collected for trace segmentation network training, using three different biological samples: ribosomes, the $\beta_2$ adrenergic receptor, and DNA Holliday junctions. For each dataset, approximately 1,500 traces were manually classified by two experts, including start and photobleaching points. Details of the experiments and system illustrations were provided in our previous study[37].

### Network training

Our model consists of an embedding layer, a BiLSTM, a Conditional Random Field (CRF), and an HMM layer, arranged sequentially as illustrated in Fig. 1(a). The input two-color fluorescence trace is first processed through the model, and frame-wise labels are predicted by the CRF layer. Based on these predicted labels, we extract segments annotated as donor-photobleached or acceptor-photobleached and feed them into the HMM layer to infer discrete hidden states. After state assignment, we analyze the state transition patterns within these segments—for example, transitions from state 0 to state 1—and identify representative patterns characteristic of the experimental condition.

In stage one, the neural network was constructed using the PyTorch framework and executed on a single NVIDIA 3090 Ti GPU. The architecture includes a bi-LSTM-CRF layer, as illustrated in Figure 1(a), consisting of three bi-LSTM layers with a hidden dimension of 32, followed by a CRF layer. During training, a dropout rate of 0.2 was applied to prevent overfitting, with a learning rate set to 0.005 and a batch size of 128. In stage two, the Hidden Markov Model (HMM) was implemented using the hmmlearn framework, employing a Gaussian mixture model with two components and covariance fitting restricted to diagonal elements to minimize overfitting. This dual-stage approach ensures robust handling of both sequential data and probabilistic state transitions within the dynamic dataset.

**Application of DASH to identify events**

For full FRET sequences $X = (X_1, X_2, \ldots, X_m)$, where each sequence $X_i = (x_{i,1}, x_{i,2}, \ldots x_{i,t})$ consists of two-color fluorescence signals, the bi-LSTM-CRF layer is designed to predict the segmentation label $\hat{y}_{i,t}$ for $x_{i,t}$. Here, $\hat{y}_{i,t}$ represents a label from the label set $\mathcal{L}$. The final classification includes three types: {*donor photobleached, acceptor photobleached, discarded*}. The label set $\mathcal{L}$ follows the BIO (Begin-Intermediate-Omit) format commonly used in Named Entity Recognition (NER) literature and is defined as $\mathcal{L} = \{O, db_b, db_i, db_e, ab_b, ab_i, ab_e\}$, where 'db' stands for donor photobleached and 'ab' stands for acceptor photobleached. The 'O' label represents the discarded type and is omitted from further analysis. The suffixes 'b,' 'i,' and 'e' denote the beginning, intermediate, and end of a labeled segment, respectively.

After bi-LSTM layer, we get embeddings $h_{i,t}$ for signals $x_{i,t}$, and CRF layer then applies a linear transformation to compute the emission scores for each label:

$$E_{i,t} = W h_{i,t} + b$$

The CRF layer defines a structured probability distribution over possible label sequences using emission scores $E_{i,t}$ and transition scores $A_{\hat{y},y}$, which represents the transition matrix for moving from label $\hat{y}$ to label y. The score of a sequence $\hat{Y}$ given an input $X$ is:

$$S(X_i, Y_i) = \sum_{t=1}^{T} E_{t,y_{(i,t)}} + \sum_{t}^{T} A_{y_{(i,t-1)}, y_{(i,t)}}$$

Where the first term sums the emission scores of the chosen labels, the second term sums the transition scores of consecutive labels. The model computes the partition function $Z(X_i)$, summing over all possible label sequences $\hat{Y}_i$:

$$Z(X_i) = \sum_{\hat{Y}_i} S(X_i, \hat{Y}_i)$$

The loss function is negative log-likelihood of the correct sequence $Y_i$,

$$\ell_{CRF} = -\sum_{i}^{N} \log \frac{\exp(S(X_i, Y_i))}{Z(X_i)}$$

In contrast to prior work employing CNN-LSTM-based models, DASH utilizes a bi-LSTM-CRF architecture, where the CRF layer optimizes transitions between different labels when generating the classification result. In terms of loss function, bi-LSTM optimizes categorical cross-entropy for individual labels, whereas bi-LSTM-CRF maximizes the likelihood of the entire label sequence by considering both emission and transition probabilities. Although bi-LSTM is computationally more efficient, bi-LSTM-CRF introduces additional complexity due to sequence-level normalization and decoding but significantly enhances accuracy in structured sequence labeling tasks. For example, the model avoids outputting sequences where a "db_i" (donor photobleached, intermediate) label precedes a "db_b" (donor photobleached, beginning) label, thereby enhancing the accuracy of sequential classification and the reliability of trace interpretation.

$$\ell_{bi-LSTM} = -\sum_{i}^{N} \sum_{t}^{T} \log P(y_{(i,t)} \mid x_{(i,t)})$$

In stage 2, the task becomes an unsupervised task since we don't have the labels. We define a subset of sequence elements where the predicted label is not "O". The final output for stage 2 is a discrete state $z_{i,t}$ assigned to $s_{i,t}$.

$$s_{i,t} = \{ x_{(i,t)} \mid \hat{y}_{i,t} \neq o \}$$

The HMM is defined by the following, transition probability matrix $A$, which is the transition probability from state $k$ to $k'$, and emission probability matrix $B$, which is the emission probability matrix of observing $s_{i,t}$ given state $z_{i,t}$.

$$A_{kk'} = P(z_{i,t} = k' | z_{i,t-1} = k)$$
$$B_k(S_{i,t}) = P(s_{i,t} | z_{i,t} = k)$$

To estimate the hidden states $Z$, the target is to compute the forward probability, which can be computed recursively.

$$a_{i,t}(k) = P(s_{i,1}, s_{i,2}, \ldots, s_{i,t}, z_{i,t} = k)$$

$$a_{i,t}(k) = \left[ \sum_{k'} a_{i,t-1}(k') A_{k'k} \right] B_k(s_{i,t})$$

Viterbi algorithm is used to find the most likely state sequence $Z$,

$$\delta_{i,t}(k) = \max_{k'} [\delta_{i,t-1}(k') A_{k'k}] B_k(s_{i,t})$$

The optimal state at each step is,

$$z_{i,t}^* = \arg \max_k \delta_{i,t}(k)$$

The advantage of using Conditional Random Fields (CRF) layers in the biLSTM-CRF architecture, compared to CNN+LSTM models, lies in their ability to model dependencies between output labels explicitly. CRFs are particularly effective for structured prediction tasks, such as sequence labeling, because they consider the relationships between neighboring labels, ensuring globally optimal label assignment. This is crucial in tasks like trace classification in single-molecule fluorescence experiments, where the transitions between states (e.g., donor photobleached, acceptor photobleached) are interdependent and follow specific patterns. While CNN+LSTM architectures focus on feature extraction and temporal dependencies, they lack mechanisms to enforce label consistency across the sequence.

## Statistics and reproducibility

All statistical analyses were conducted using the Origin 2021 (OriginLab Corporation) software or *PYTHON*. Three repeats were performed to ensure the consistency and reliability of the data. No data were excluded from the analyses, and the sample sizes were not predetermined using statistical methods.

## References

1. Stryer, L. & Haugland, R. P. Energy transfer: a spectroscopic ruler. *Proc Natl Acad Sci U S A* **58**, 719–26 (1967).

2. Ha, T. *et al.* Probing the interaction between two single molecules: Fluorescence resonance energy transfer between a single donor and a single acceptor. *Proc Natl Acad Sci U S A* **93**, 6264–8 (1996).

3. Hellenkamp, B. *et al.* Precision and accuracy of single-molecule FRET measurements—a multi-laboratory benchmark study. *Nat Methods* **15**, 669–676 (2018).

4. Lerner, E. *et al.* Toward dynamic structural biology: Two decades of single-molecule Förster resonance energy transfer. *Science* **359**, eaan1133 (2018).

5. Rundlet, E. J. *et al.* Structural basis of early translocation events on the ribosome. *Nature* **595**, 741–745 (2021).

6. Blanchard, S. C., Gonzalez, R. L., Kim, H. D., Chu, S. & Puglisi, J. D. tRNA selection and kinetic proofreading in translation. *Nat Struct Mol Biol* **11**, 1008–14 (2004).

7. Rundlet, E. J. *et al.* Structural basis of early translocation events on the ribosome. *Nature* **595**, 741–745 (2021).

8. Ferguson, A. *et al.* Functional dynamics within the human ribosome regulate the rate of active protein synthesis. *Mol Cell* **60**, 475–86 (2015).

9.  Nishima, W. *et al.* Hyper-swivel head domain motions are required for complete mRNA-TRNA translocation and ribosome resetting. *Nucleic Acids Res* **50**, 8302–8320 (2022).

10. Newton, M. D. *et al.* DNA stretching induces Cas9 off-target activity. *Nat Struct Mol Biol* **26**, 185–192 (2019).

11. Globyte, V., Lee, S. H., Bae, T., Kim, J. & Joo, C. CRISPR /Cas9 searches for a protospacer adjacent motif by lateral diffusion. *EMBO J* **38**, e99466. (2019).

12. Zeng, Y. *et al.* The initiation, propagation and dynamics of CRISPR-SpyCas9 R-loop complex. *Nucleic Acids Res* **46**, 350–361 (2018).

13. Yang, M. *et al.* Nonspecific interactions between SpCas9 and dsDNA sites located downstream of the PAM mediate facilitated diffusion to accelerate target search. *Chem Sci* **12**, 12776–12784 (2021).

14. Akyuz, N. *et al.* Transport domain unlocking sets the uptake rate of an aspartate transporter. *Nature* **518**, 68–73 (2015).

15. Das, D. K. *et al.* Direct Visualization of the Conformational Dynamics of Single Influenza Hemagglutinin Trimers. *Cell* **174**, 926–937 (2018).

16. Asher, W. B. *et al.* Single-molecule FRET imaging of GPCR dimers in living cells. *Nat Methods* **18**, 397–405 (2021).

17. Sakon, J. J. & Weninger, K. R. Detecting the conformation of individual proteins in live cells. *Nat Methods* **7**, 203–5 (2010).

18. Zhao, J. *et al.* Ligand efficacy modulates conformational dynamics of the μ-opioid receptor. *Nature* **629**, 474–480 (2024).

19. Ngo, T. T. M. *et al.* Effects of cytosine modifications on DNA flexibility and nucleosome mechanical stability. *Nat Commun* **7**, 10813 (2016).

20. Manz, C. *et al.* Single-molecule FRET reveals the energy landscape of the full-length SAM-I riboswitch. *Nat Chem Biol* **13**, 1172–1178 (2017).

21. Niu, X. *et al.* Structural and dynamic mechanisms for coupled folding and tRNA recognition of a translational T-box riboswitch. *Nat Commun* **14**, 7394 (2023).

22. Cole, F. *et al.* Super-resolved FRET and co-tracking in pMINFLUX. *Nat Photonics* **18**, (2024).

23. McKinney, S. A., Joo, C. & Ha, T. Analysis of single-molecule FRET trajectories using hidden Markov modeling. *Biophys J* **91**, (2006).

24. Li, C. B. & Komatsuzaki, T. Aggregated markov model using time series of single molecule dwell times with minimum excessive information. *Phys Rev Lett* **111**, (2013).

25. Sgouralis, I. *et al.* A Bayesian Nonparametric Approach to Single Molecule Förster Resonance Energy Transfer. *Journal of Physical Chemistry B* **123**, (2019).

26. Schmid, S., Götz, M. & Hugel, T. Single-Molecule Analysis beyond Dwell Times: Demonstration and Assessment in and out of Equilibrium. *Biophys J* **111**, (2016).

27. Li, J., Zhang, L., Johnson-Buck, A. & Walter, N. G. Automatic classification and segmentation of single-molecule fluorescence time traces with deep learning. *Nat Commun* **11**, 5833 (2020).

28. Wanninger, S. *et al.* Deep-LASI: deep-learning assisted, single-molecule imaging analysis of multi-color DNA origami structures. *Nat Commun* **14**, 6564 (2023).

29. de Lannoy, C. V., Filius, M., Kim, S. H., Joo, C. & de Ridder, D. FRETboard: Semisupervised classification of FRET traces. *Biophys J* **120**, (2021).

30. Krizhevsky, A., Sutskever, I. & Hinton, G. E. ImageNet classification with deep convolutional neural networks. *Commun ACM* **60**, (2017).

31. Van Houdt, G., Mosquera, C. & Nápoles, G. A review on the long short-term memory model. *Artif Intell Rev* **53**, (2020).

32.     Chen, C. *et al.* Single-Molecule Fluorescence Measurements of Ribosomal Translocation Dynamics. *Mol Cell* **42**, 367–377 (2011).

33.     Heng, J. *et al.* Function and dynamics of the intrinsically disordered carboxyl terminus of β2 adrenergic receptor. *Nat Commun* **14**, 2005 (2023).

34.     Zhang, Y. *et al.* General strategy to improve the photon budget of thiol-conjugated cyanine dyes. *J Am Chem Soc* **145**, 4187–4198 (2023).

35.     Sun, R., Zhao, Y., Wang, W., Liu, J. J. G. & Chen, C. Nonspecific interactions between Cas12a and dsDNA located downstream of the PAM mediate target search and assist AsCas12a for DNA cleavage. *Chem Sci* **14**, 3839–3851 (2023).

36.     Xing, W., Li, D., Wang, W., Liu, J.-J. G. & Chen, C. Conformational dynamics of CasX (Cas12e) in mediating DNA cleavage revealed by single-molecule FRET. *Nucleic Acids Res* **52**, 9014–9027 (2024).

37.     Zhou, S. *et al.* Deep learning based local feature classification to automatically identify single molecule fluorescence events. *Commun Biol* **7**, 1404 (2024).

38.     Liu, J. J. *et al.* CasX enzymes comprise a distinct family of RNA-guided genome editors. *Nature* **566**, (2019).

39.     Burstein, D. *et al.* New CRISPR-Cas systems from uncultivated microbes. *Nature* **542**, (2017).

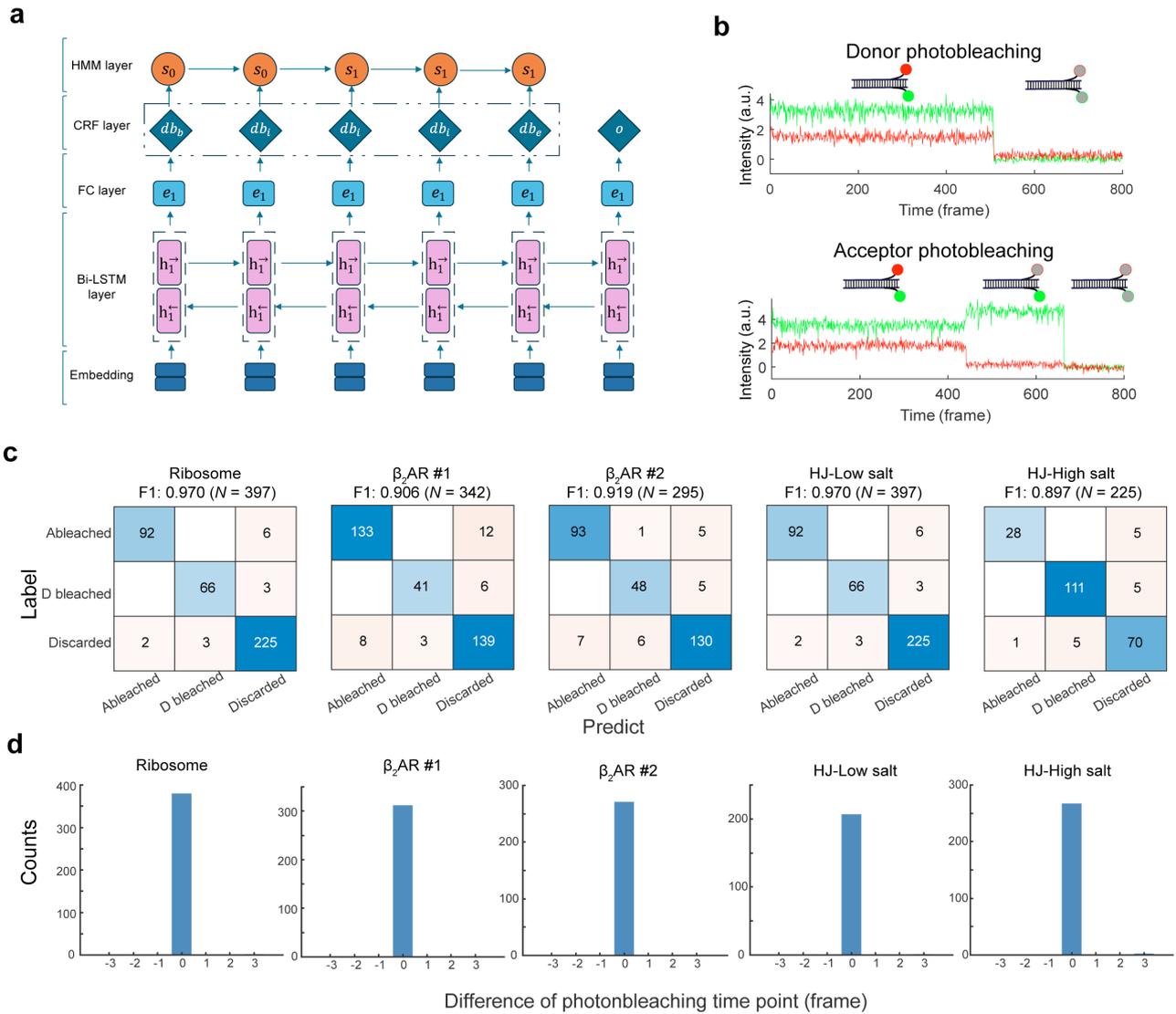

**Figure 1.** Workflow of trace classification performance of DASH. (a) DASH consists of two stages: trace classification and state assignment. Bi-LSTM refers to Bidirectional Long Short-Term Memory, while FC layer represents the Fully Connected layer. CRF stands for Conditional Random Field, and HMM represents Hidden Markov Model. (b) Examples of traces classified as 'donor photobleaching' (top) and 'acceptor photobleaching' (bottom) are shown. 'Donor photobleaching' is characterized by the simultaneous disappearance of both donor and acceptor, whereas 'acceptor photobleaching' is defined by the acceptor disappearing first, followed by the donor. Green represents the donor, while red indicates the acceptor. The black lines depict double-strand DNA. (c) Confusion matrices illustrate the differences between predicted classifications and labeled data. D bleached refers to donor photobleaching, and A bleached indicates acceptor photobleaching. $N$ represents the number of traces, and F1 is the harmonic mean of precision and recall. (d) The difference in photobleaching time points between predictions and labels is presented, with an exposure time of 100 ms per frame.

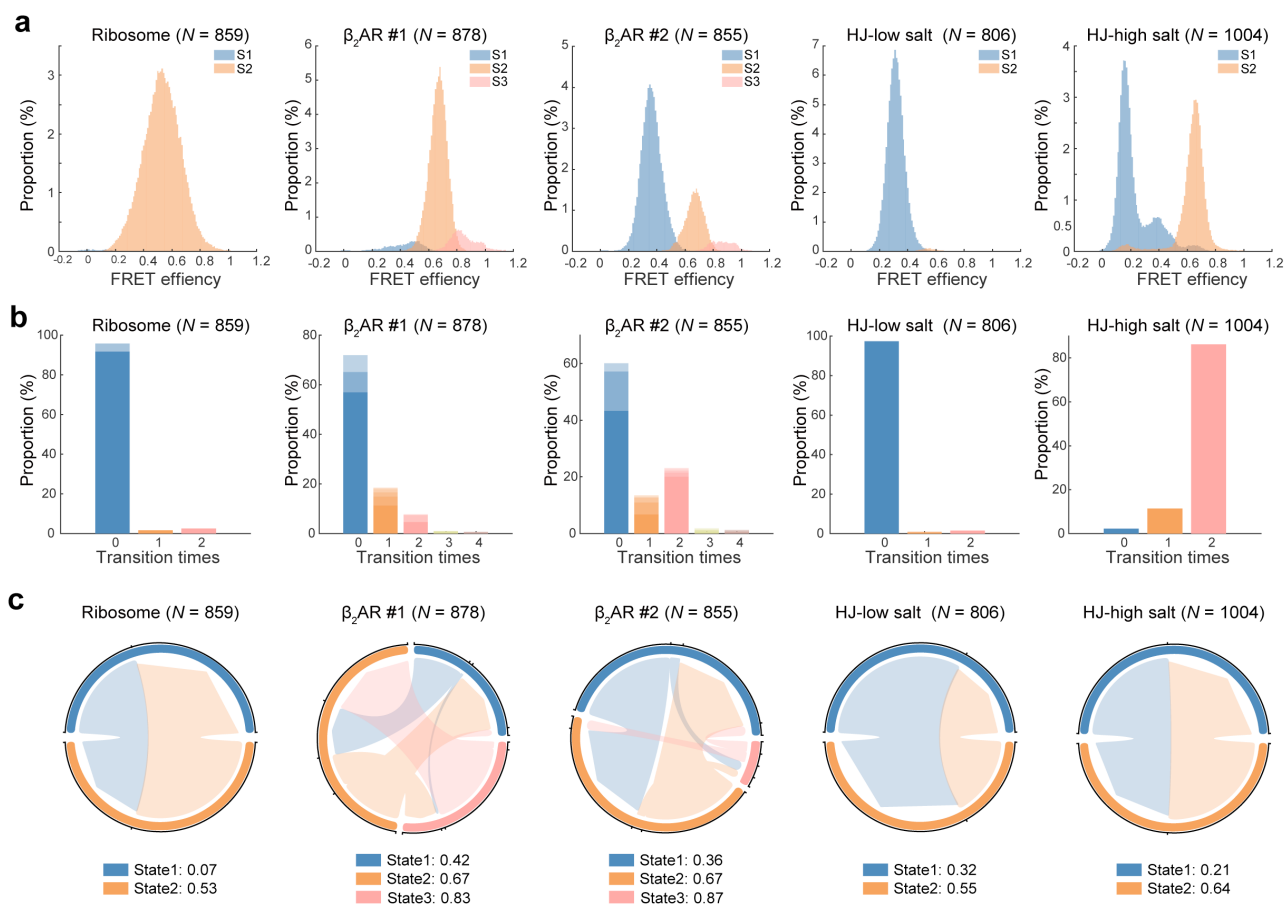

**Figure 2.** Analysis of automatic sorting results on static two-color traces according to DASH (a) FRET distributions are extracted from traces that have been classified and assigned states by DASH. (b) Distributions of transition times in a fluorescent trace determined by DASH. Different shades of the same color represent various patterns. (c) State-to-state transition direction and probabilities of transient traces. $N$ represents the number of traces analyzed.

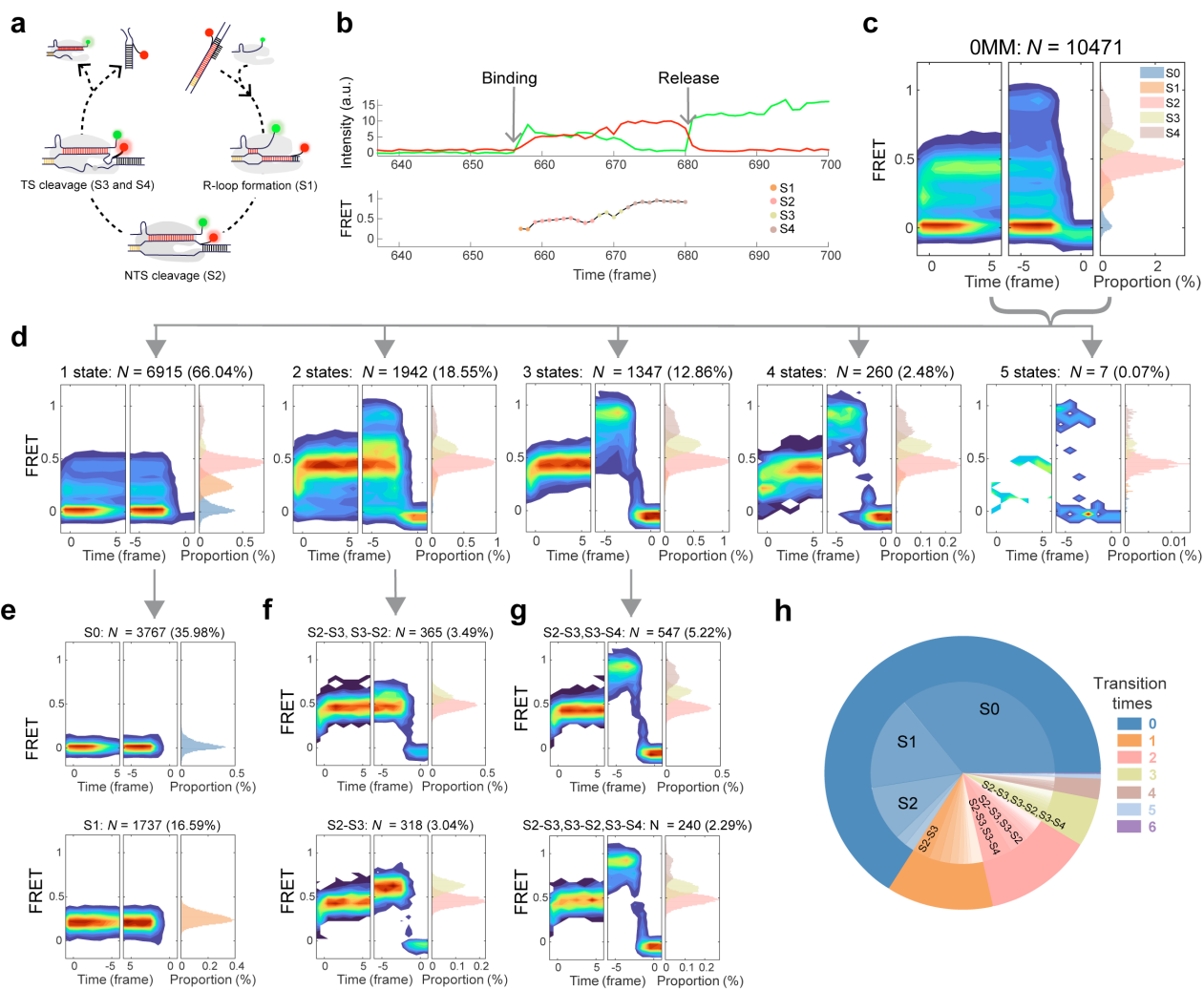

**Figure 3.** Automatic sorting based on DASH of AsCas12a with fully matched target DNA. (a) Illustrations and a typical single-molecule fluorescence trace depict the corresponding states of Cas12a during DNA cleavage. Cy3 (green dot) is attached to the 3' end of crRNA, while Cy5 (red dot) is linked to the target DNA strand, positioned 28 nucleotides downstream of the PAM sequence. State 1 (S1) corresponds to the complex with initial R-loop formation, State 2 (S2) represents the non-target strand (NTS) cleavage state, and States 3 and 4 (S3 and S4) correspond to the target strand (TS) cleavage states. (b) Representative single-molecule FRET trajectories of AsCas12a cleavage are shown, with four distinct FRET states indicated and assigned as S1 to S4, ranging from low to high values. The green line represents the donor fluorophore (Cy3), while the red line represents the acceptor fluorophore (Cy5). (c) Time-dependent FRET probability density plots are presented, synchronized at the appearance of FRET (t = 0, left panel) or at the disappearance of FRET (t = 0, middle panel). These plots allow for careful examination of FRET changes in the Cas12a/crRNA/DNA complex after formation or prior to the disassembly of the ternary complex, respectively. (d) Time-dependent FRET probability density plot based on automatic sorting of the number of states present in an event. (e-g) Time-dependent FRET probability density plot based of certain pattern appearing in an event within 1 state (e), 2 states (f), and 3 states (g) categories, please refer to Supplementary Table 7 for complete sorting results based on DASH. (h) A double pie chart illustrates transient trace transition times, with different shades of the same color representing the same number of transitions but differing patterns. The exposure time for each frame is 500 ms. $N$ represents the number of traces analyzed.

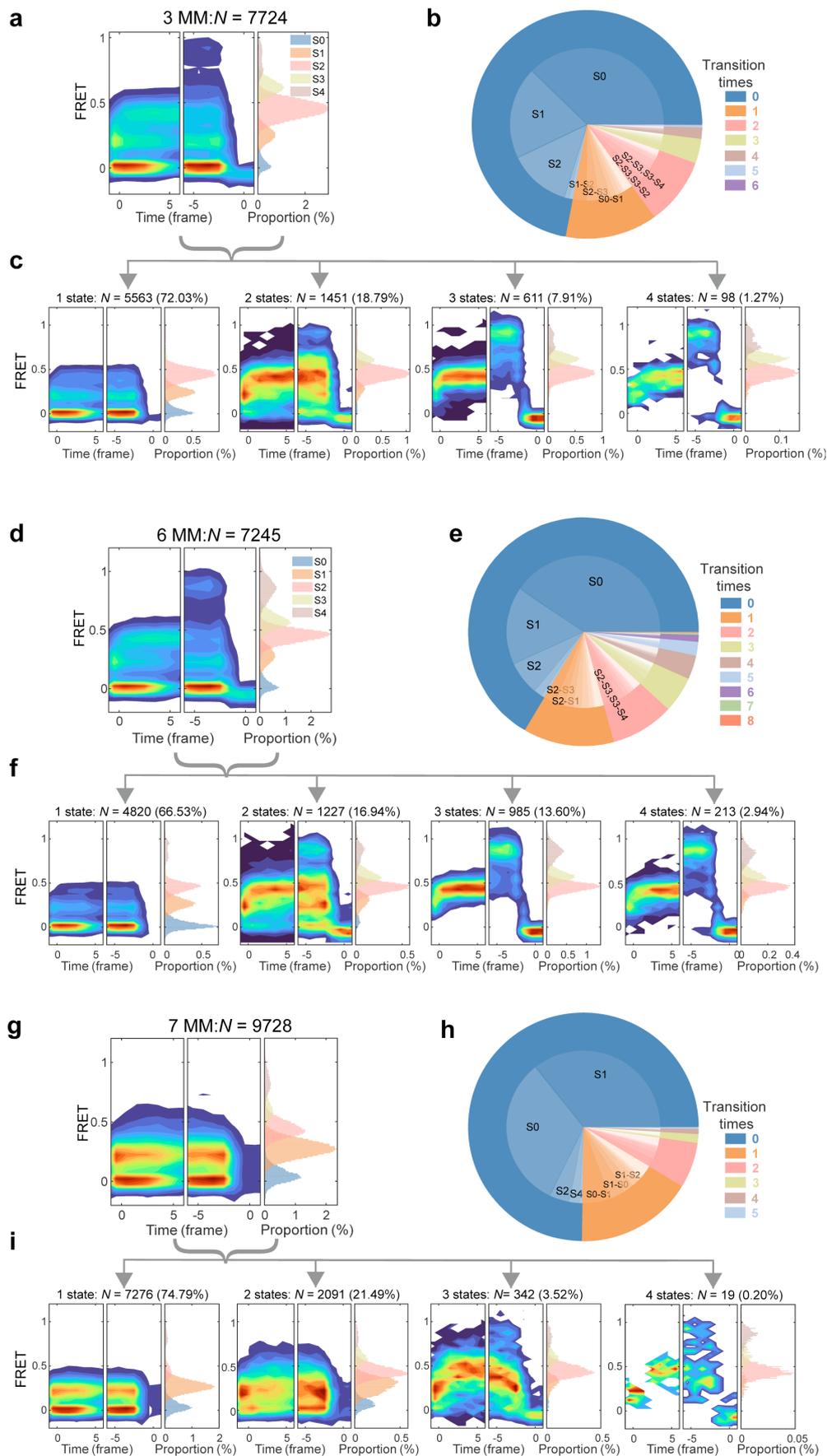

**Figure 4.** Automatic sorting based on DASH of AsCas12a with mismatch target DNA. (a-c) Time-dependent FRET probability density plots of all fluorescent events (a), automatic sorting by number of states present in an event (c), and a double pie chart illustrates transient trace transition times (b) of AsCas12a with 3MM. (d-f) Time-dependent FRET probability density plots of all fluorescent events (d), automatic sorting by number of states present in an event (f), and a double pie chart illustrates transient trace transition times (e) of AsCas12a with 6MM. (g-i)Time-dependent FRET probability density plots of all fluorescent events (g), automatic sorting by number of states present in an event (h), and a double pie chart illustrates transient trace transition times (i) of AsCas12a with 7MM. In pie chart, different shades of the same color represent the times of transitions is the same but the pattern is different. 3 MM, 6 MM, and 7 MM stand for DNA targets containing 3, 6, and 7 PAM-distal mismatches, respectively, towards crRNA at their PAM-distal ends. Please refer to Supplementary Tables 7-9 for complete sorting results based on DASH. The exposure time for each frame is 500 ms. $N$ represents the number of traces analyzed.

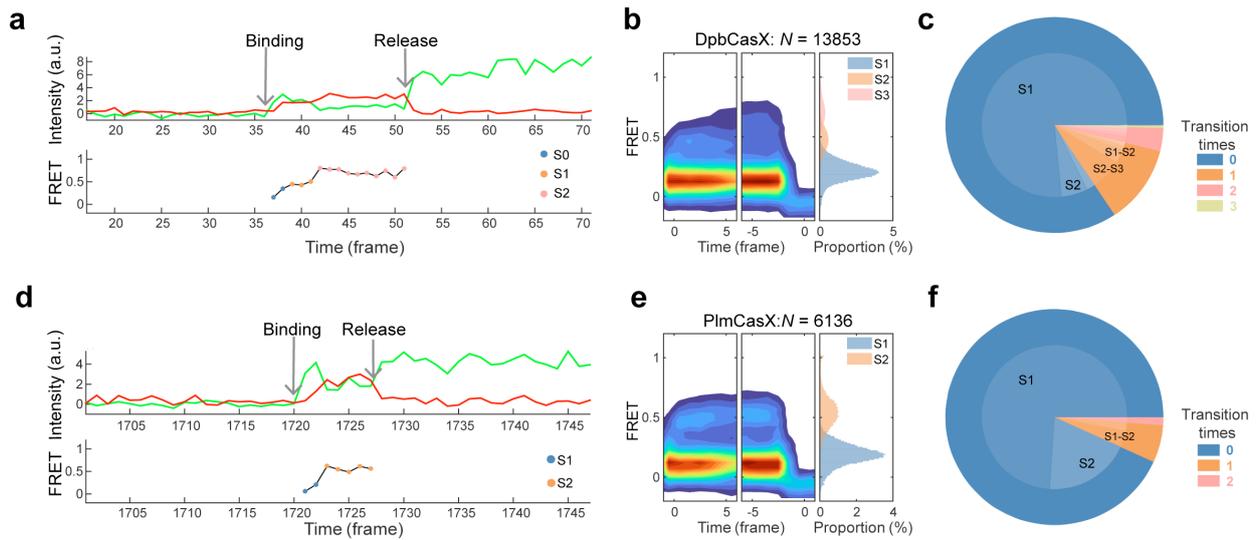

**Figure 5.** Automatic sorting based on DASH of CasX with fully matched target DNA. (a) Representative single-molecule FRET trajectories of dpbCasX cleavage are presented, with three distinct FRET states indicated and assigned as S1 to S3, ranging from low to high values. The green line represents the donor fluorophore (Cy3), while the red line represents the acceptor fluorophore (Cy5). (b) Time-dependent FRET probability density plots for all fluorescent events of dpbCasX are shown. (c) A double pie chart illustrates the transient trace transition times of dpbCasX. (d) Representative single-molecule FRET trajectories of plmCasX cleavage are displayed, with two distinct FRET states indicated and assigned as S1 to S2, from low to high values. (e) Time-dependent FRET probability density plots for all fluorescent events of plmCasX are provided. (f) A double pie chart shows the transient trace transition times of plmCasX. The exposure time for each frame is 500 ms. *N* represents the number of traces analyzed.